\documentclass[twocolumn,letterpaper,secnumarabic,amssymb, nobibnotes, aps, prl]{revtex4-1}        
\usepackage{epsfig}
\usepackage{dcolumn}
\usepackage{epstopdf}
\usepackage{graphicx}
\begin{document}

\title{Quantum body in uniform electric fields}  

 \author{S. Selenu}
\affiliation{}

\begin{abstract}
\noindent  The advent in this century of nano and microelectronics  requires, by part of physicists and engineers, the need of an explanation of electrical  phenomena such as the interaction of a body with external electric fields at its atomic level,i.e. where is the case of the appearance of  the quantum nature of phenomena involved in the projecting of nano an micro devices. It is mainly focused in this article the problem of calculating eigenstates of quantum matter interacting with a uniform external electric field, whose solution is reported via a DFT(Density Functional Theory) variational approach\cite{Kohn,Kohn1}.
\end{abstract}

\date{\today} 
\maketitle

 \section{Introduction}
Nowaday the approach many scientists gave to the projecting or either experimenting of quantum nano and micro electronic devices\cite{micro,nano} put an effort on searching of solutions of the quantum eigenstates of matter due to the scaling of the length of the dimensions of involved electrical phenomena. An Hamiltonian\cite{Lan1,Lan2,Lan3} approach for the theoretical solving of the problem then requires the writing of eigenvalue differential equations can be solved trough modern computational techniques we shall include, in the solving of the problem, with the aim to simplify the projecting procedure itself. Several works were brought forward by many authors\cite{Umi,Pascqui,Resta,Vanderbilt} not yet giving a clear DFT based\cite{Kohn,Kohn1} eigenvalue equation, here in this paper found by mean of a variational approach extending the Kohn-Sham differential equations. In the next section it will be introduced several problems arising from the solving of the quantum equations due mainly to the search of solutions in the context of infinite periodic systems or either of a finite sampling of matter made by finite or   periodic replicas. In the last part of the article will be reported conclusions.
\noindent  

\section{Search of quantum eigenstates from a variational principle}
\label{hg}
Kohn-Sham approach nowaday understood being a DFT approach routinely employed in order calculating eigenstates of quantum matter when is the case of matter itself being in its minimum of the energy, will be used towards this article introducing a new differential equation related to the interaction of a quantum body  with an external uniform electric field $\bf{E_0}$\cite{Jackson,Pauli,Nenciu}. In the physics of crystals or either when bulk calculations of quantum dielectric properties of matter are concerned, enforces the author of this paper, to abandon the idea of an external electrostatic  potential:

\begin{eqnarray}
\label{pot}
V_{\bf{E_0}}=-e{\bf{E_0}} \cdot \bf{r}
\end{eqnarray}

being $e$ the electron charge as also apparently unbounded, due to the repetition in space of a unit cell, building an infinite periodic crystal. A Fourier approach is employed in order to avoid the  unboundedness by calculating the quantum energy of the system trough a Fourier series of the position operator $\bf{r}$ . The suggested approach allows then exploiting  symmetries of the system while introducing  clearly  main contributions to the energy as given by  the as so forth believed unbounded potential. Firstly it is put on relevance the quantum energy density should have the following expression:

\begin{eqnarray}
\label{Energy}
\langle \Psi | H | \Psi \rangle=\frac{1}{V} [\int \Psi^* \frac{-\hbar^2 \nabla^2}{2m} \Psi + \int e\rho V - \int \rho e\bf{E_0}\cdot \bf{r}]
\end{eqnarray}

extending then Hohenberg and Kohn problem. Here wave functions are normalized with respect to the volume of the system has been explicitly written on eq.(\ref{Energy}). Recasting equation $\ref{Energy}$ in terms of the interaction of the electric field with the dipole of the system it can be written:

\begin{eqnarray}
\label{Energy2}
VE= \int \Psi^* \frac{-\hbar^2 \nabla^2}{2m} \Psi + \int \rho V +E_D\\\nonumber 
E_D=-e\bf{E_0}\cdot \int \rho  \bf{r}
\end{eqnarray}

Several authors\cite{Umi,Pascqui,Resta,Vanderbilt} do approach the problem not having yet recognized the quantum dipole in an infinite crystal being periodic while recognizing it being a Berry phase\cite{Vanderbilt} of quantum eigenstates, written in terms of the linear connection\cite{Resta}. The latter form of the quantum dipole is strictly periodic and boundary free\cite{Selenudipolo,Resta,Vanderbilt} but does not allow for a clear DFT approach in order writing  of an eigenvalue differential equation by functional derivatives with respect to the electronic charge  $\rho$, the latter considered the electronic charge distribution of the quantum matter interacting with the electric field. A Fourier analysis of the periodicity of  the quantum dipole instead shown in infinite crystals put on relevance  the symmetry of the electronic charge density of the system has,  it to allow recognizing directly  its boundedness. It will be firstly considered a finite sample of matter while analizing the dielectric response of a quantum body then extend the reasoning to an infinite crystal where periodicity of a unit cell is asked in order quantify the bulk polarization\cite{Linesglass}it may arise by the reaction of the crystal eigenstates to the external electric field. Also, it can always be divided the quantum body in several unitary virtual cells of the reference space coordinates to each associated a smooth quantum wave function part of the quantal eigenstate of matter $\Psi$it will be called $\Psi_{\bf{R}}$ recognizing $\bf{R}$ being a set of  lattice vectors identifying  the unitary virtual cells. The quantum energy reads:

\begin{eqnarray}
\label{Energy3}
VE= \sum_{\bf{R}}\int \Psi^*_{\bf{R}} \frac{-\hbar^2 \nabla^2}{2m} \Psi_{\bf{R}} + \sum_{\bf{R}}\int e\rho_{\bf{R}} V_{\bf{R}} + E_D\\\nonumber E_D=-e\bf{E_0}\cdot \sum_{\bf{R}}\int \rho_{\bf{R}}  \bf{r}
\end{eqnarray}

being $\rho_{\bf{R}}$ the part of the smooth electronic charge density contained in a unitary virtual cell as well as for the electrostatic potential $V_{\bf{R}}$ displaced by the first unitary cell by a lattice vector $\bf{R}$. When integrations are referred to the virtual cells  the dummy $\bf{r}$ coordinate in the integrals allows writing  the part of the energy term involving the quantum dipole as it follows:

\begin{eqnarray}
\label{Energy4}
E_D=-e\bf{E_0}\cdot \sum_{\bf{R}}\int \rho_{\bf{R}} \bf{r}-e\bf{E_0}\cdot \sum_{\bf{R}}\int \rho_{\bf{R}} \bf{R}
\end{eqnarray}

where integrations are over a mesh grid of points referred to the first virtual  cell of the  lattice construction whose contribution is left equal to zero by symmetry in an infinite crystal as the lattice vectors sum up to zero as it will be directly shown by direct evaluation the last term of eq.(\ref{Energy4}), by considering each cell of a crystal having the same amount of electronic charge stored in it. An electric energy term $E_D$ arise from the interaction of the dipole with the external electric field where it is evident,  dividing it due in two contributions,i.e a bulk contribution and a lattice contribution. The bulk contribution strictly depends on the charge density in the interior of the virtual cells while the lattice contribution is due to the charge density at the boundaries. As we shall see the whole sum of the energy is invariant with the choice of the virtual lattice unless symmetries on the system are exploited as in the case of an infinite periodic crystal, while making each part arbitrary. Nonetheless we are mainly focus on the calculation of the invariant sum, considering now evident the energy being well bounded and amenable of  being variated with respect to wave functions on the virtual cell. Before performing a variational analysis  of the bounded energy we can evaluate the dipole moment density via Fourier analisys and also calculating the energy density it to be directly variated. Let us consider firstly:

\begin{eqnarray}
\label{Energy5}
{\bf{P}}=\frac{e \sum^N_{\bf{R}}\int [\rho  \bf{r}]}{NV}
\end{eqnarray}
 
being $V$ the volume of the virtual unit cells chosen for the evaluation of the microscopic dipole per cell, it a formula  can be used during the writing of this paper showing itself being very useful when employed in the calculation of dipole moment of matter in finite systems of either in infinite crystals. Variating the energy functional $E$  reported in eq.(\ref{Energy3}) with respect to the $\Psi_{\bf{R}}$ wave functions we can calculate the macroscopic dipole of the finite sample of matter, as it follows:

\begin{eqnarray}
\label{Energy6}
\frac{\delta E}{\delta \Psi^*_{\bf{R}}}= \frac{\delta }{\delta \Psi^*_{\bf{R}}} (\sum_{\bf{R}}\int \Psi^*_{\bf{R}} \frac{-\hbar^2 \nabla^2}{2m} \Psi_{\bf{R}} + \sum_{\bf{R}}\int e\rho_{\bf{R}} V_{\bf{R}} +E_D) =0\\\nonumber
\end{eqnarray}

The result of this performed variation can be then employed to any encountered problems of calculating the response of matter to external electrostatic fields whose search of a set of quantum eigenstates is reaced by recognizing it having at hands the whole set of differential equations whose solutions are  the parts of the smooth wave function $\Psi_{\bf{R}}$  stored in the virtual cell building the total wave function $\Psi$:

\begin{eqnarray}
\label{Energy7}
H_{\bf{R}}\Psi_{\bf{R}}= \frac{-\hbar^2 \nabla^2}{2m} \Psi_{\bf{R}} + V_{\bf{R}} \Psi_{\bf{R}} -e{\bf{E_0}}\cdot  {[\bf{r+R}]}\Psi_{\bf{R}}\\\nonumber
\end{eqnarray}

being the position operator $\bf{r}$ varying in the first virtual  unit cell of the virtual lattice. It becomes evident that when a symmetric array of unitary virtual cells is employed in order to calculate the wave functions $\Psi_{\bf{R}}$ the lattice contribution to the dipole in eq.(\ref{Energy4}) becomes vanishing when it is the case of having a constant electronic charge in the virtual unit cell. A direct example is recognized being in a finite bulk crystal where the virtual unit cell is a crystal lattice unit cell having a constant electronic charge on each cell. The same behaviour is encountered in first principle calculations when ab initio model schemes\cite{Martin} are employed in order simulate quantum electronic or either atomic structure of matter. Equation  $\ref{Energy5}$  of the quantum dipole reduces to:

\begin{eqnarray}
\label{Energy8}
{\bf{P}}=\frac{e}{V} \int [\rho  \bf{r}]
\end{eqnarray}
we can directly evaluate via Fourier analysis of the electronic charge density and of the position $\bf{r}$ restricted to the unit cell.  Expanding in Fourier series the position operator and the charge density of the second equality in eq.(\ref{Energy4}) in terms of Fourier components can being either an expansion in plane wave \cite{Martin} of  reciprocal vectors $\bf{G}$ it easy understood the expression of the dipole moment of the electronic charge reconducible to an appealing calculable form in computational  physics and an electrical energy term as so given:

\begin{eqnarray}
\label{Energy9}
E_D=-e\bf{E_0}\cdot  \sum_{\bf{G}} \rho_{\bf{G}}  \bf{r}_{\bf{G}}
\end{eqnarray}

leaving to a bulk dipole density (dipole moment per unit volume ) equal to:

\begin{eqnarray}
\label{Energy10}
{\bf{P}}= \frac{e}{V} \sum_{\bf{G}} \rho_{\bf{G}}  \bf{r}_{\bf{G}}
\end{eqnarray}

invariant with respect the choice of the unit cell and amenable of direct calculation by first principle techniques where the set of eigen differential equations reduce to a unique differential equation on the unit cell written as:

\begin{eqnarray}
\label{Energy11}
H\Psi= \frac{-\hbar^2 \nabla^2}{2m} \Psi+ V \Psi -e{\bf{E_0}}\cdot {[\bf{r}]}\Psi
\end{eqnarray}

being, as stated the position operator $ [\bf{r}]$, only varying in the first unit cell of the crystal lattice, makig the useful its employement in electronic structure calcualtions. In the next part of the article will be reported conclusions.

\section{Conclusions}
This article put on clearness a long avoided problem, encountered by condensed matter physicists, on the explanation of the interaction of quantum matter explicitly state being the electronic charge density. Today routinely measured on laboratories and simulated via DFT computational schemes, the employed electronic charge density so obtained will help to  calculate  eigenstates of matter of the a so long believed unbounded energy problem via the reported  set of eigenvalue differential equations can be helpful for the calculations of quantum eigenstates itself interacting with electrostatic fields and a formula of the polarization reported in eq.(\ref{Energy5}) it reducing to formula eq. (\ref{Energy10}) in the case of crystal bulk calculations.

\end{document}